\documentstyle[twoside,fleqn,espcrc2,epsf]{article}

\title{$K \rightarrow \pi \pi$ Decays with Domain Wall Fermions:
Towards Physical Results}
 
\author{Robert D. Mawhinney\address{Department of Physics, Columbia
  University, New York, NY, 10027, USA}\thanks{Supported in part by
  the US Department of Energy and the RIKEN-BNL Research Center.
  This work was done in collaboration with T. Blum, N. Christ,
  C. Cristian, C. Dawson, G. Fleming, X. Liao, G. Liu, S. Ohta,
  A. Soni, P. Vranas, M. Wingate, L. Wu and Y. Zhestkov.  Presented
  at Lattice '00, Bangalore, India} \\
  RBC Collaboration
}
 
\begin{document}

\def\thepage{CU--TP--992}
\thispagestyle{myheadings}

\begin{abstract}

We are using domain wall fermions to study $K \rightarrow \pi \pi$
matrix elements by measuring $K \rightarrow \pi$ and $K \rightarrow 0$
matrix elements on the lattice and employing chiral perturbation theory
to relate these to the desired physical result.  The residual chiral
symmetry breaking of domain wall fermions with a finite
extent in the fifth dimension impacts these measurements.  Using the
Ward-Takahashi identities, we investigate residual chiral symmetry
breaking effects for divergent quantities and study pathologies of the
quenched approximation for small quark mass.  We then discuss the
$\Delta S = 1$ operator $O_2$, where chiral symmetry is vital for the
subtraction of unphysical effects.

\end{abstract}

\maketitle

\section{INTRODUCTION}

Lattice QCD allows a first principles determination of hadronic
properties, including the matrix elements of operators between hadronic
states.  Among the more interesting matrix elements are those of the
$\Delta S = 1 $, four-fermion operators which are responsible
for the decay of a kaon into two pions.  These matrix elements are
needed to relate the values of the Cabbibo-Kobayashi-Maskawa matrix in
the standard model to physical measurements.  Direct measurements of $K
\rightarrow \pi \pi$ on the lattice are not possible
\cite{maiani-testa}, due to the two pions in the final state, although
a recent suggestion involving tuning the lattice volume
\cite{luscher-lellouch} looks promising.  Alternatively, lowest order
chiral perturbation theory can be used to relate $K \rightarrow \pi
\pi$ to $ K \rightarrow \pi$ and $K \rightarrow 0$ matrix elements
\cite{bernard}, which can be measured on the lattice.  This approach
has been extensively studied and, while quite straight-forward in
principle, requires careful control over a wide range of phenomena.

Of central importance to a calculation relying on chiral perturbation
theory is precise control over chiral symmetry on the lattice.  A major
advance in this long-standing problem is the idea of four-dimensional
chiral fermions arising from mass defects in a higher-dimensional
fermion formulation \cite{kaplan}.  With improved chiral symmetry at
finite lattice spacing, the lattice Ward-Takahashi identities, without
complicated renormalization, are much closer to the continuum ones.
Here we seek to understand how the residual chiral symmetry breaking
effects for domain wall fermions enter the removal of unphysical,
power divergent contributions, which is needed to get $K \rightarrow
\pi \pi $ from $K \rightarrow \pi$ and $K \rightarrow 0 $ matrix
elements.

\section{DOMAIN WALL FERMIONS}

The RIKEN-BNL Research Center/BNL/ Columbia (RBC) Collaboration has
recently finished extensive quenched simulations of the hadron spectrum
and low-energy QCD physics, using domain wall
fermions \cite{shamir}, with a particular focus on the residual
chiral symmetry breaking effects due to finite values for the fifth
dimension, $L_s$ \cite{dwf_hadron_00}.  (Similar work was also reported
in \cite{cppacs}.)  This work produced quantitative
values for the residual quark mass $m_{\rm res}$, which enters low
energy QCD physics through the effective quark mass, $m_{\rm eff} = m_f
+ m_{\rm res}$, where $m_f$ is the input bare quark mass.  The residual
quark mass was shown to be small, $\sim 1/30$ the strange quark mass
for $\beta = 6.0$ and $L_s = 16$.

It was also found that topological near-zero modes of the domain wall
fermion operator, which are not suppressed by a fermionic determinant
in quenched simulations, have pronounced effects on quark propagators
for small $m_f$.  In addition, for large volumes where topological
near-zero mode effects are suppressed, a non-linear dependence of
$m_{\pi}^2$ on $m_f$ was found, consistent with a quenched chiral
logarithm.  Thus, even though the chiral properties of domain wall
fermions are quite improved over other formulations, these quenched
pathologies can appear in simulation results.

We have also finished measuring $K \rightarrow \pi$ and $K \rightarrow
0$ matrix elements for the 10 $\Delta S = 1$ effective four-fermion
operators \cite{ciuchini} relevant to determinations of $\epsilon'/
\epsilon$ and the real parts of the $\Delta I = 1/2 $ and 3/2
amplitudes for $K \rightarrow \pi \pi $ decays.  We have results for
200 quenched lattices of size $16 \times 32$ with $L_s = 16$ and $\beta
= 6.0$ and a variety of light and heavy quark masses.  (For more
details about the simulations see \cite{blum}.)  The analysis of these
results requires control over the residual chiral symmetry breaking
effects and an understanding of the role of quenched pathologies in our
data.  This report is a summary of our progress to date.

A final step in determining physical matrix elements is the matching
of continuum and lattice operators.  This has been done using
non-perturbative renormalization \cite{npr} and details can be
found in \cite{cdawson}.

\pagenumbering{arabic}
\addtocounter{page}{1}

\section{WARD IDENTITY CHECK}

Furman and Shamir \cite{shamir} defined axial transformations
for domain wall fermions, which lead to the Ward-Takahashi identity
\begin{eqnarray}
  \Delta_\mu \langle {\cal A}^a_\mu(x) O(y) \rangle  & =  &
	2m_f \langle J^a_5(x) O(y) \rangle  \label{eq:ward_tak_id} \\
    & +  & 2 \langle J^a_{5q}(x) O(y) 
	\rangle + i \langle \delta^a O(y) \rangle \nonumber.
\end{eqnarray}
Here ${\cal A}^a_\mu(x)$ at the four-dimensional point $x$ includes
spinors from all values of the the fifth coordinate.  The pseudoscalar
density $J^a_5(x)$ is constructed from four-dimensional fields in the
standard way and $J^a_{5q}(x)$ is constructed from ``mid-point''
spinors at $s = L_s/2 -1$ and $L_s/2$.  (The notation is as in
\cite{dwf_hadron_00}.)

For physics that involves momentum scales much less than the cutoff,
the extra term in Eq.\ \ref{eq:ward_tak_id} involving $J^a_{5q}$ (the
``mid-point'' term), gives a contribution of the form $ J^a_{5q} =
m_{\rm res} J^a_5 $.  This defines $m_{\rm res}$ to $O(a^2)$, since
the low-energy QCD physics of domain wall fermions should by
describable by an effective Lagrangian.  However, in matrix elements
involving divergent quantities one only expects that $ J^a_{5q} \sim
O(m_{\rm res}) J^a_5$.  Since divergent expressions enter in $K
\rightarrow \pi $ amplitudes and we must manipulate the Ward-Takahashi
identities to remove them, these effects need to be understood.

To test our understanding of $m_{\rm res}$ effects and our data, we
first consider $O(y) = J^a_5(y)$ in Eq. \ref{eq:ward_tak_id}.
Summing over $x$ gives (no sum on $a$)
\begin{equation}
  \sum_x \langle [ m_f \,J^a_5(x) + J^a_{5q}(x) ] J^a_5(y) \rangle
    - \langle \overline{q} q \rangle = 0
  \label{eq:int_wi}
\end{equation}
The term in the sum without $J^b_{5q}$ gives $m_f/m_\pi^2$ pion pole
terms plus $ O(m_f/a^2)$ contact terms along with constant terms and
terms of higher order in $m_f$.  Similarly the mid-point term gives
$m_{\rm res}/m_\pi^2$  pion pole terms, plus $O(m_{\rm res}/a^2)$
contact terms as well as constant contributions.  The pion pole terms
enter as $(m_f + m_{\rm res})/m_\pi^2$ in Eq.\ \ref{eq:int_wi} while
the contact contributions enter as $O(m_f) + O(m_{\rm res})$.  This
form for the pion pole term is consistent with $m_\pi = 0$ at
$m_f = -m_{\rm res}$.

Thus we are lead to expect a Gell-Mann--Oakes--Renner relation
for domain wall fermions with the form
\begin{equation}
  \frac{m_\pi^2 f^2}{2(m_f + m_{\rm res})} =
    - \langle \bar{q}q \rangle_{m_f}
    + O(\frac{m_f}{a^2})
    + O(\frac{m_{\rm res}}{a^2})
  \label{eq:gmor}
\end{equation}
This relation holds at non-zero values for $m_f$, with the $O(m_f/a^2)$
contact term contribution being cancelled by the similar term in $
\langle \overline{q}q \rangle$.  Of course, there may be other quenched
pathologies entering this expression, such as effects from topological
near-zero modes or quenched chiral logarithms.

\begin{figure}[tb]
\epsfxsize=\hsize
\epsfbox{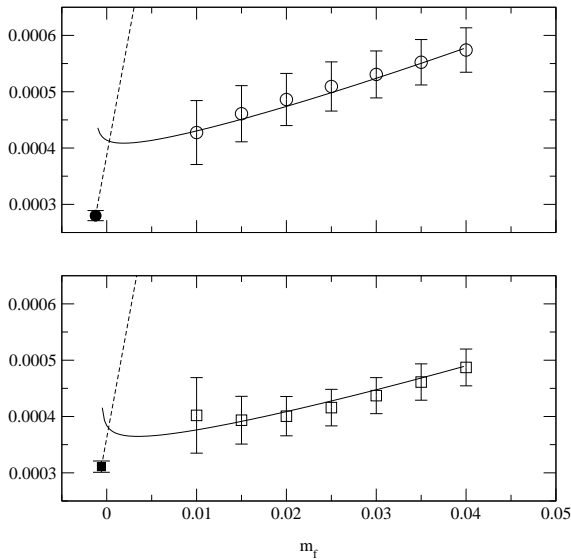}
\caption{The GMOR relation for $L_s = 16$ (upper panel) and
$L_s =24$ (lower panel).  The open symbols are the left side
of Eq.\ \ref{eq:gmor} and the filled symbols are
$ -\langle \overline{q}q \rangle$.  The dashed line gives the
$m_f/a^2$ dependence of $ -\langle \overline{q}q \rangle$ and the
solid line includes the effects of quenched chiral logarithms in
$m_\pi^2$.}
\label{fig:gmor}
\end{figure}

We have measured $m_\pi$, $f$, $m_{\rm res}$ and $ -\langle
\overline{q}q \rangle$ for $16^3 \times 32$ lattices at $\beta = 6.0$
\cite{dwf_hadron_00} and plot our results in Figure \ref{fig:gmor}.
One sees that for $L_s = 16$, the larger mass points are approximately
linear, but do not seem to extrapolate towards the value of $ -\langle
\overline{q}q \rangle$ at $m_f = - m_{\rm res}$.  For $L_s = 24$, the
extrapolation is much closer, as is expected since $m_{\rm res}$ has
decreased by almost a factor of 2.  Also shown are solid lines
representing the left-hand side of Eq.\ \ref{eq:gmor} when a quenched
chiral logarithm fit is used for $m_\pi^2(m_f)$.  One sees that the
effect of chiral logarithms is to exacerbate the disagreement in the
extrapolation.  Thus we have evidence for residual mass contact terms
in this simple Ward-Takahashi identity.

In conclusion, we note that for $L_s = 16$, the quantity $m_\pi^2 f^2 /
(- 2 \langle \overline{q}q \rangle (m_f + m_{\rm res}))$, which is 1 in
chiral perturbation theory, differs from 1 by around 20\% for our
data.  Thus one should avoid introducing extraneous powers of this
quantity while manipulating the data.  We now turn to the
Ward-Takahashi identities useful for subtracting the divergent
contributions to $K \rightarrow \pi$ matrix elements.

\section{TOWARDS $\langle \pi^+ \pi^- | O_n | K^0 \rangle$ }

Here we discuss our tests of a particular method for
removing the divergent effects in $K \rightarrow \pi $ matrix elements.
Following \cite{sharpe}, we define a subtracted operator
\begin{equation}
  \tilde{O}_n = O_n + \eta_n [ ( m_d + m_s ) \bar{s} d +
    (m_d - m_s ) \bar{s} \gamma_5 d ]
  \label{eq:osub}
\end{equation}
where $O_n$ is a $\Delta S = 1$ four-fermion operator and
$\eta_n$ is chosen such that
\begin{equation}
  \langle 0 | \tilde{O}_n | K^0 \rangle = 0.
\end{equation}
Then following \cite{bernard}, but noting that
contact terms in the Ward-Takahashi identities coming from $m_{\rm res}$
effects will enter, we find
\begin{equation}
  \langle \pi^+ \pi^- | O_n | K^0 \rangle = 
  \frac{i(m_{K^0}^2 - m_{\pi^+}^2)} {f} \frac{d \langle
  \pi^+ | \tilde{O}_n | K^+ \rangle }{ d m_M^2}
\end{equation}
where the derivative removes any contact terms proportional to
$m_{\rm res}$.

\begin{figure}[htb]
\epsfxsize=\hsize
\epsfbox{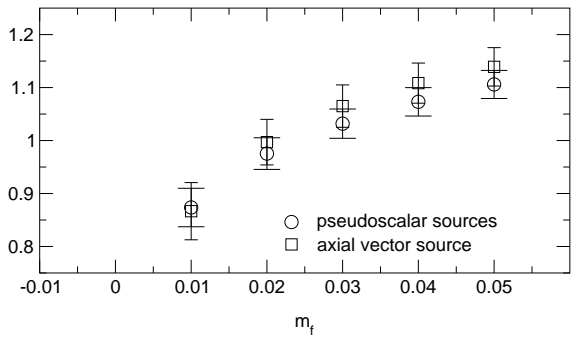}
\caption{The left-hand side of Eq. \ref{eq:k2pi_sbard} plotted
versus quark mass.}
\label{fig:k2pi_sbard}
\end{figure}

The successful removal of the effects of mixing with the dimension
three operators implemented in Eq.\ \ref{eq:osub} requires chiral
symmetry.  We first check this with a simpler Ward-Takahashi identity, obtained
by inserting an operator $O = \overline{s}(y)(1 - \gamma_5)
d(y)  \; \bar{u}(z) \gamma_5 s(z)$ in Eq.\ \ref{eq:ward_tak_id}.
The pion pole terms in the Ward-Takahashi identity then lead to
\begin{equation}
  \begin{array}{c}
      \langle \pi | \bar{s} d | K \rangle
       \frac{2(m_f + m_{\rm res})}{m^2_\pi} \hfill  \\
      \;\;\;\;\;\; = 1 + O(m_f) + O(m_{\rm res}) + \cdots
  \end{array}
  \label{eq:k2pi_sbard}
\end{equation}
There are no power divergent operators in this Ward-Takahashi identity,
so the coefficient of the $m_f$ and $m_{\rm res}$ contact terms are
finite, but there may still be quenched chiral logarithm effects
present.  In Figure \ref{fig:k2pi_sbard} we plot the left-hand side of
Eq.\ \ref{eq:k2pi_sbard} versus $m_f$.  We see that for larger $m_f$,
the points appear to extrapolate reasonably well to 1, but there is
noticeable curvature for small quark masses.  The figure includes
results for axial vector sources, which agree with those for
pseudoscalar sources, indicating that topological near-zero modes are
not the origin of this non-linearity.  We are investigating whether
this can be explained as a quenched chiral logarithm.

\begin{figure}[htb]
\epsfxsize=\hsize
\epsfbox{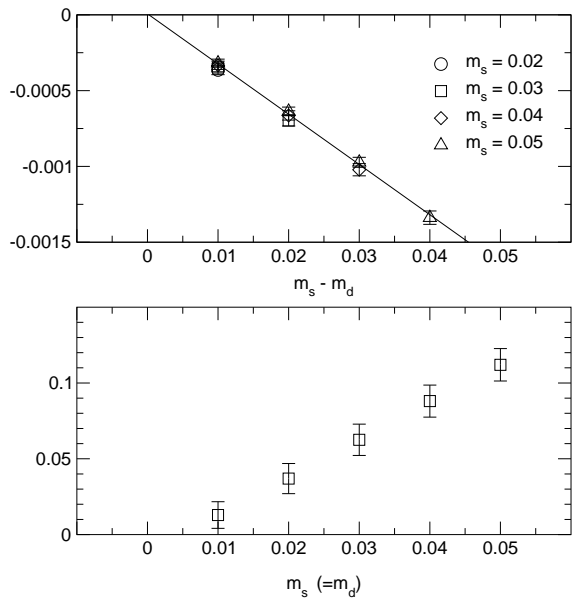}
\caption{The value for $\langle 0 | O_2 | K^0 \rangle / \langle 0 |
\bar{s}\gamma_5 d | K^0 \rangle$ is shown in the upper panel.
The lower panel is a plot of the $K \rightarrow \pi$ matrix element
of $\tilde{O}_2^{\frac{1}{2}}$ normalized as in Eq.\ \ref{eq:O2_sub}.}
\label{fig:O2}
\end{figure}

We determine $\eta_n$ from measurements of $\langle 0 | O_n| K^0
\rangle$ for non-degenerate masses for the strange and down quarks.
The upper panel of Figure \ref{fig:O2} shows a plot of $\langle0 | O_2|
K^0 \rangle / \langle 0 | \bar{s}\gamma_5 d | K^0 \rangle$ versus $m_s
- m_d$.  We see good linearity which allows for a reliable
determination of $\eta_2$, with statistically errors at the few percent
level for this particular operator.

With the value of $\eta_2$ determined, $\tilde{O}_2$ is known and
Figure \ref{fig:O2} shows
\begin{equation}
  \frac{ \langle \pi^+ | O_2^{\frac{1}{2}} + \eta_2(m_s + m_d)
    \bar{s} d | K^+ \rangle}
    { \langle \pi | P | 0 \rangle \langle 0 | P | K \rangle}
  \label{eq:O2_sub}.
\end{equation}
where $O_2^{\frac{1}{2}}$ is the $\Delta I = 1/2$ part of $O_2$.  In
lowest order chiral perturbation theory, the denominator is a constant,
so the desired $K \rightarrow \pi \pi$ matrix element is given by the
slope and the intercept is related to $m_{\rm res}$.  We are looking
carefully for non-linearities in our data, paying particular attention
to the quenched chiral logarithm effects recently calculated in
\cite{golterman}.

\section{CONCLUSIONS}

We have seen that the Ward-Takahashi identities for domain wall
fermions include $O(m_{\rm res}/a^2)$ contact terms which can be
removed by taking derivatives with respect to $m_f$.  Our data also
show evidence for non-linear dependence on $m_f$, some of which may be
due to quenched chiral logarithms.  We have shown that the subtracted
operator $\tilde{O}_2$ can be determined and are now doing the
subtractions for all the operators.

The calculations reported here were run on the QCDSP computers
at Columbia University and the RIKEN-BNL Research Center.

\end{document}